\documentclass[aps,preprint,showpacs,showkeys]{revtex4}
\usepackage{amssymb,amsmath,epsf,graphicx}
\usepackage[ansinew]{inputenc}
 
\begin{document}
\title{A criterion for population inversion by arbitrary pulses}
\author{Werner Jakubetz}
\affiliation{Fakultät für Chemie, Universität Wien, Währinger Strasse 17, A-1090 Wien, Austria}
\author{Christoph Uiberacker}
\email{christoph.uiberacker@unileoben.ac.at}
\affiliation{Institut für Physik, Montanuniversität Leoben, Franz-Josef Strasse 18, 8700 Leoben, Austria}

\begin{abstract}
A theoretical investigation of population dynamics in an $N$-level system driven by pulses of arbitrary shape is presented, which is based on Floquet theory. The pulse only has to have finite support in time, which is well fulfilled in experiments. Furthermore the dynamics must be approximately time-local, which restricts the applicability to negligible memory effects. The rotating-wave approximation is not invoked. We discuss control of both the real and imaginary part or just the modulus of any component of the final wavefunction. A necessary criterion on the Floquet quasienergies of pulses of arbitrary shape is obtained for population inversion in the case $N=2$, which is a generalization of the $\pi$-pulse criterion previously demonstrated by Holthaus et.~al. We find that only the eigenvectors of the propagator contain enough information to give a sufficient criterion. By analysing the propagator in an $N$-level system we estimate the number of real parameters of the pulse that have to be adjusted to control transitions between arbitrary initial and final states, including superposition states. The propagator can be recast into block diagonal form, leading to a description by an effective 2-level-system composed of the initial and final state, for which our criterion is still valid, however additional conditions on the pulse are required. In all cases the number of conditions on the pulse parameters increases linearly with $N$.
\end{abstract}
\maketitle
 
\newpage 

\section{Introduction}

For a long time there have been attempts to control specific quantum transitions with high efficieny using pulses that might itself consist of a train of subpulses \cite{C_contol}. While a number of approaches rest on combining experiment and simulation using feedback \cite{C_contol_exp}, in the present paper we study model systems in the most general setup to find simple criteria of control. 
There are many applications where such criteria would be desirable, as in charge excitation and transport in biological systems \cite{C_bio}, constructing efficient QBits for quantum computers \cite{C_qbits}, or spin injection and magnetic transport in spintronic devices \cite{C_spintronic}. 

Not even the two-level system (2LS) can be solved analytically for the general case. Traditional approximations are the rotating wave approximation (RWA) \cite{C_rwa} and Floquet theory \cite{C_Floquet}. The former is justified only for small detunings between field- and transition-frequency while the latter in its original form is applicable for strictly period fields (where it is exact), although in adiabatic approximation it can be extended to sufficiently slowly varying pulse envelopes \cite{C_adiabApprox}.

In recent years short pulses of only few optical cycles \cite{C_exp_shortpulses} have become available, which are not suitable for adiabatic approximations. For such pulses semiclassical strong field theory \cite{StrongFld} applies, which however is valid only in the limit of large quantum numbers. 
Therefore most theoretical studies have used numerical techniques, including the relatively new topic of dynamics in molecular systems under the influence of such strong pulses \cite{C_molecule_pulse}. Recently we contributed an approximate treatment of $N$-level systems in the presence of such pulses \cite{C_uiberacker09}.

Regarding full population inversion, Holthaus and Just have previously used Floquet theory to analyse population transfer in a pulse with adiabatic envelope within the RWA \cite{C_holthaus_prl92,C_holthaus_pra94}. In the present paper we show analytically that a generalization of that criterion provides a necessary condition for population inversion under {\it arbitrary} pulses of finite support under fairly general conditions, i.e. without invoking RWA (retaining however the semiclassical dipole approximation). An important additional observation is that a sufficient condition for population inversion is provided only by the Floquet eigenvectors.
We generalize the discussion by investigating either full control of the wavefunction (populations and phases) or just the populations. The former case is important for interference experiments. 
We evaluate the number of conditions on real parameters of the pulse, noting that the nonlinear equations for the pulse parameters resulting from the Floquet equations might have no physical solution.

We add a detail concerning notation. Consider the setup of arbitrary orthogonal initial and final vectors in an $N$-level system.
Whenever the square modulus of each component of the fina wavefunction (at the end of the pulse) is equal to the corresponding quantity of a given final target vector, we say that the corresponding pulse achieves (generalized) ''population inversion'' (PI); the case where initial and final vector are both system eigenstates corresponds to the usual notion of a resonance. If the wavefunction itself is equal to the final vector, we use the term ''phase selective population inversion'' (PSPI).  

On the other hand in case initial and final vector are not orthogonal we deal with a general control scenario and use the terms ''population control'' (PC) and ''phase selective population control'' (PSPC). As this scenario goes somewhat beyond the topic of population inversion, we 
discuss it separately in appendix \ref{App_nonortho}.

\section{Theory}

In the following we consider the interaction of a quantum system, approximated by $N$ levels, with a laser pulse in terms of the semiclassical dipole approximation, noting that the derivations do not depend on the details of the interaction. Setting $\hbar=1$ and choosing a basis $\mathbf{B}_0:=\{\mathbf b^{(1)},\dots,\mathbf b^{(N)}\}$, the time-dependent Schrödinger equation becomes
\begin{equation} 
	i\partial_t \psi_j = \sum_{k} H_{jk}\psi_k \quad , \quad H_{jk} = \epsilon_j\delta_{jk} - \mu_{jk}E(t) \quad . 
	\label{schreq}
\end{equation}
Here $\delta_{jk}$ stands for the Kronecker symbol.
The $\psi_j$, $j=1,\dots,N$, are the time dependent expansion coefficients of the state described by the wavefunction $\psi$, the $\epsilon_j$ are the eigenvalues, and the $H_{jk}$ are elements of the Hamiltonian $\mathbf{H}$. $E(t)$ is the projection of the field onto the dipole operator with expectation values $\mu_{jk}$. The diagonal elements $\mu_{jj}$ represent permanent dipole moments. Note that our derivation also applies to non-hermitean Hamiltonians, which are used in studies of open systems \cite{C_opensys}, as long as memory effects are negligible during the pulse. 

In order to use the Floquet method we consider the experimental pulse $E(t)$ of duration $T$ to represent one period of an auxilary periodic field $E_p(t)$. By the assumed time-local behaviour populations at time $T$ will be equal for both fields.

\subsection{Floquet Theory}

We define the propagator transporting the basis $\mathbf{B}(t_0)=\mathbf{B}_0$ as $\mathbf{B}(t):= \mathbf{U}(t,t_0)\mathbf{B}(t_0)$ and rewrite the Schrödinger equation as
\begin{equation}
	i\mathbf{G}(t) \mathbf{U}(t,t_0) = 0 \quad , \quad G(t) := \partial_t + i\mathbf{H}(t) \quad ,
	\label{parallel_trans}
\end{equation} 
using the contravariant operator $\mathbf{G}$ with its local representation $\mathbf{G}(t)$ along the one-dimensional (1d) submanifold of propagation in the unitary group $\rm U(N)$.
After completing a closed loop in $t$ we obtain $\mathbf{B}(T+t_0)=\mathbf{U}(T+t_0,t_0)\mathbf{B}(t_0)$. In mathematics $\mathbf{U}(T+t_0,t_0)$ is called the monodromy matrix. Note that when transporting along the loop for a second time we start with $\mathbf{B}(T+t_0)$. Due to the periodicity of the auxilary field $E_p(t)$ we have $\mathbf{G}(t)=\mathbf{G}(T+t)$, the propagator is equal to the one of the first loop, and $\mathbf{B}(2T+t_0)=[\mathbf{U}(T+t_0,t_0)]^2\mathbf{B}(t_0)$. We conclude that the set of all loops forms an abelian group \cite{C_homotopy}, which then has only 1d irreducible representations \cite{C_groups}. In this way the following exponential homomorphism is obtained,
\begin{equation}
	\mathbf{U}(nT+t_0,t_0) = \exp(in\mathbf{\Omega}T) \quad , \quad \Omega_{jk} = \omega_j\delta_{jk} \quad .
	\label{E_prop_floquet}
\end{equation} 
The $\omega_j$ are the well-known Floquet quasienergies, which are real whenever $\mathbf{H}=\mathbf{H}^{\dagger}$ and complex otherwise.

\subsection{Population inversion in two-level systems}

In order to find criteria for PI it suffices to analyse the general form of the propagator matrix after a full period. We set $t_0=0$ for the time when the external perturbation is switched on, so that $\mathbf{U}(0,0)=\mathbf{I}_2$, the 2d unit matrix (below, $\mathbf{I}_n$ will denote the $n\times n$ unit matrix). Furthermore we abbreviate $\mathbf{V}:=\mathbf{U}(T,0)$. The propagator of an $N$-level system belongs to the manifold $\rm U(N)$. For $N=2$ we expect $4$ parameters to determine the propagator \cite{C_groups}. The equations $\mathbf{V}\mathbf{V}^{\dagger}=\mathbf{V}^{\dagger}\mathbf{V}=\mathbf{I}_2$ lead to
\begin{equation}
	\mathbf{V} = \left[
			\begin{array}{cc}
				e^{i\gamma_a}\cos(\delta) & e^{i\gamma_b}\sin(\delta) \\
				-e^{i\gamma_c}\sin(\delta) & e^{i\gamma_d}\cos(\delta) \\
			\end{array} \right]
			\quad , \quad \gamma_a,\gamma_b,\gamma_c,\gamma_d,\delta\in\mathbb{R}
\end{equation}
with the additional relation $\gamma_a+\gamma_d=\gamma_b+\gamma_c+2j\pi$, $j\in\mathbb{Z}$.
Defining the quantities  
\begin{equation}
	 \chi:=\frac{\gamma_a+\gamma_d}{2} \quad , \quad \Delta_1:=\frac{\gamma_a-\gamma_d}{2} 
	 		\quad , \quad \Delta_2:=\frac{\gamma_b-\gamma_c}{2}
\end{equation}
and extracting a phase factor we finally arrive at
\begin{equation}
	\mathbf{V} = \exp(i\chi)\left[
			\begin{array}{cc}
				e^{i\Delta_1}\cos(\delta) & e^{i\Delta_2}\sin(\delta) \\
				-e^{-i\Delta_2}\sin(\delta) & e^{-i\Delta_1}\cos(\delta) \\
			\end{array} \right]  \quad .
	\label{E_Prop}
\end{equation}
The matrix $\exp(-i\chi)\mathbf{V}$ has unit determinant and hence is an element of ${\rm SU}(2)$. Only the parameter $\delta$ determines the modulus of elements of $\mathbf{V}$, whereas all other parameters just contribute phases.

Using a relation for the Wronski determinant from the theory of ordinary differential equations \cite{C_sanchez79}, namely
\begin{equation}
	\det \mathbf{U}(t,0) = \det \mathbf{U}(0,0) \exp\left\{-i\int_{0}^t {\rm tr}~\mathbf{H}(t') dt'\right\}
		\quad ,
\end{equation}
we can immediately identify
\begin{equation}
	 \chi = -\frac{1}{2}\int_0^T {\rm tr}~\mathbf{H}(t') dt'
	 \label{E_chi}
\end{equation}

We proceed with the setup of PI investigated in \cite{C_holthaus_prl92} and take as the initial state
$\psi^{(\rm i)} =[1,0]^{\rm T}$ (a superscript ${\rm T}$ denotes the transpose, not to be confused with the period $T$) with population only in level i, and use $\psi^{(\rm f)} =[0,1]^{\rm T}$ with population only in level f.
In order to find situations more readily realizable in experiment we concentrate on the case $n=1$ ($n$ the number of pulses) and {\it assume} that such a pulse exists (as suggested by numerical results, a rigorous proof of existence for the general case is yet to be given).
From the structure of the propagator in eq.~(\ref{E_Prop}) and $\psi^{(\rm i)}$ we obtain $|\psi^{(\rm f)} (T)|^2=\sin^2\delta$ for the population in f, reflecting the fact that with regard to PI only $\delta$ is relevant. PI then demands
\begin{equation}
	\delta^{\rm PI} = \frac{2k+1}{2}\pi \quad , \quad k\in\mathbb{Z} \quad .
\end{equation}
We note that $2k+1$ can be interpreted as the number of PIs during the pulse.

In order to find a necessary criterion based on the Floquet quasienergies we diagonalize $\mathbf{V}$, that is, we solve the polynomial
\begin{equation}
	\det(\mathbf{V}-\zeta\mathbf{I}) = 0 \quad ,
\end{equation}
resulting in
\begin{equation}
	\zeta_{1,2} = \frac{1}{2}\exp(i\chi)\left\{ \cos(\delta)\cos(\Delta_1) \pm 
		\left[\cos^2(\delta)\cos^2(\Delta_1) - 4\exp(-i\chi)\right]^{1/2} \right\} \quad .
	\label{E_evals}
\end{equation}
In case of PI we find that the two values $\zeta_{1,2}$ are symmetric with respect to $0$, namely
\begin{equation}
	\zeta_{1,2}^{\rm PI} = \pm i\exp(i\frac{\chi}{2}) \quad .
	\label{E_evals_res}
\end{equation}
This gives the necessary condition
\begin{equation}
	(\omega_{2}^{\rm PI}-\omega_{1}^{\rm PI})T = (2n+1)\pi \quad , \quad n\in\mathbb{Z} \quad ,
	\label{E_ResFloquetPhases}
\end{equation}
as a criterion for arbitrary pulses, analogous to the one based on an adiabatic Floquet approach in refs. \cite{C_holthaus_prl92} and \cite{C_holthaus_pra94}.
Clearly eq.~(\ref{E_ResFloquetPhases}) corresponds to an avoided crossing situation because without the off-diagonal contributions in $\mathbf{V}$ the eigenvalues would both be zero. 

With a view on applications of condition (\ref{E_ResFloquetPhases}), we note that in a pragmatic sense at first sight it does not seem to lead beyond the results obtained by integrating the Schrödinger equation numerically. However, efficient expansions exist for the Floquet determinant, which could be used to find the quasienergies up to the required approximation from a time-independent system of equations \cite{C_Floquet}. Moreover an analytic approach might answer the question of {\it existence} of a pulse effecting full PI.   

Returning to eq.~(\ref{E_evals}), we note that due to the vanishing trace of $\mathbf{V}$, $\Delta_1=\frac{2k+1}{2}\pi$ also leads to eq.~(\ref{E_ResFloquetPhases}). This shows that this criterion itself does not provide a {\it sufficient} condition indicating PI. In order to find a sufficient criterion we investigate the eigenvectors $\mathbf{e}^{1}$ and $\mathbf{e}^{2}$, which are determined from the equation
\begin{equation}
	\det(\mathbf{V}-\zeta_{\alpha}\mathbf{I})e_j^{\alpha} = 0 \quad .
\end{equation}
Using $\mathbf{e}^{1}=N_1[1,c_1]^{\rm T}$, $\mathbf{e}^{2}=N_2[c_2, 1]^{\rm T}$ we obtain the equations
\begin{equation}
	c_1(c_2^{-1}) = \frac{\zeta_{1(2)}}{\sin(\delta)}\exp(-i\Delta_2) - \exp[i(\Delta_1-\Delta_2)]\cot(\delta) \quad .
	\label{E_evec_comp}
\end{equation} 
Inserting the corresponding eigenvalue we arrive at
\begin{eqnarray}
	c_1(c_2^{-1}) & = & \frac{1}{2}\exp[-i(\Delta_2-\chi)]\Biggl\{ \cot\delta\bigl[\cos\Delta_1 - 
		2\exp\left[i(\Delta_1-\chi)\right] \bigr]\Biggr. \nonumber \\
	&& +(-) \left. \left[ \cot^2\delta\cos^2\Delta_1 
		- \frac{4\exp(-i\chi)}{\sin^2\delta} \right]^{1/2} \right\} \quad .
	\label{E_evec_comp_final}
\end{eqnarray} 
In case of PI we obtain for the rotation matrix the simple expression
\begin{equation}
	\mathbf{e}^{\rm PI} = \frac{1}{\sqrt{2}}\left[ 
		\begin{array}{cc}
			1 & -\exp(i\alpha) \\
			\exp(-i\alpha) & 1 
		\end{array} 
	\right] \quad , \quad \alpha := \Delta_2 - \frac{1}{2}(\chi+\pi) \quad ,.
	\label{E_evec_comp_res}
\end{equation} 
which clearly is unitary and provides the sufficient criterion for PI. 

We observe that if we set $\delta = \delta^{\rm{PI}}$ in eq.~(\ref{E_Prop}), the value of $\Delta_1$ is irrelevant. This is in analogy to spherical coordinates because they are a subset of the Euler angles parametrizing rotations in $\rm O(3)$, which is homomorphic to the group ${\rm SU}(2)$ determining the propagator if $\chi$ is fixed. Therefore we may interpret $\Delta_1$ as an azimuth angle while $\delta$ would represent the inclination. 
The rotation matrix $\mathbf{e}^{\rm {PI}}$ is real whenever $\Delta_2=0$ and $\chi=(2k+1)\pi$, $k\in\mathbb{Z}$. In this case $\zeta_{1,2}$ are real and $\omega_{1,2}$ is $0,\pi/T$.

In control theory the wavefunction at the end of the pulse is prescribed and the inverse problem is solved \cite{C_contol}. In order to discuss PSPC in this context, we take $\psi^{(\rm f)}=[a,b]^{\rm T}$, $a,b\in\mathbb{C}$. In case of PSPC we thus get $4$ relations describing the real and imaginary parts of $a$ and $b$,
\begin{equation}
	a = \exp[i(\chi+\Delta_1)]\cos(\delta) \quad , \quad b = -\exp[i(\chi-\Delta_2)]\sin(\delta) \quad ,
\end{equation}
of which only $3$ are independent due to $|a|^2+|b|^2=1$. Therefore PSPC determines all $3$ parameters $\Delta_1$, $\Delta_2$ and $\delta$, as expected, while $\chi$ is given by eq.~(\ref{E_chi}). On the other hand PSPI only requires to manipulate the $2$ parameters $\delta$ and $\Delta_2$ due to the fact that $\Delta_1$ is irrelevant in case $\cos(\delta)=0$.  

\subsection{$N$ levels}

First we analyse the geometrical properties of the propagator by using the Floquet representation.
Assuming the quasienergies to be real of the form $\omega_j=q_j\omega_0$, with $\omega_0:=2\pi/T$, and restricting to the first Floquet zone, $q_j\in[0,1)$, we distinguish the cases of (i) rational $q_j=n_j/m$ (with $n_j,m\in\mathbb{N}$ relative prime), and (ii) at least one of the $q_j$ is irrational. In the first case, after $m$ pulses we arrive back in the initial state. Due to the translation structure, in terms of real parameters the $m$ vectors lie on a great circle of the $2N-1$ dimensional unit sphere in $\mathbb{C}^N$. For $m$ even we have the additional inversion property that to each state $\psi$, its ''mirror'' image $-\psi$ is also among the $m$ vectors.
In the second case of irrational $q_j$ the (infinite) number of vectors densely fill the great circle. 

Control of the dynamics is determined by the equation $\psi^{(\rm f)}=\mathbf{V}\psi^{(\rm i)}$. In case of PSPC this results in $2N$ equations for the real and imaginary parts of the components of $\psi^{(\rm f)}$. Due to normalization of $\psi^{(\rm f)}$ we therefore have to fix $2N-1$ real parameters of the pulse for PSPC. 

If only the populations $|\psi^{(\rm f)}_j|^2$ are relevant and the phases do not matter (setup of PC), we simply have to use the correct number of equations which amounts to replacing $2N$ by $N$. Using again $||\psi^{(\rm f)}||^2=1$, this leads to $N-1$ conditions for population control.

As is well known, time-reversal symmetry is present if $\mathbf H(t)=\mathbf H^{\dagger}(-t)$ holds, and $\psi^{(\rm f)\dagger}:=\psi^{\dagger}(T)$ is mapped into $\psi^{(\rm i)\dagger}:=\psi^{\dagger}(0)$ via $\mathbf{U}(0,T)=\mathbf{U}^{\dagger}(T,0)$ by the equation $i\partial_t \psi^{\dagger}(-t) = \psi^{\dagger}(-t)\mathbf{H}^{\dagger}(-t)$. When the field strength of the pulse is nonzero, this symmetry is not present because usually $\mathbf{H}(t) \ne \mathbf{H}^{\dagger}(-t)$. However, due to the translation symmetry of the Hamiltonian in $T$ in the Floquet problem we have time-reversal symmetry when considering only times that are multiples of $T$, due to $\mathbf{H}(nT)=\mathbf{H}(-nT)=\mathbf{H}^{\dagger}(-nT)$ with $n\in\mathbb{N}$ if the system is not open. This symmetry does however not provide extra information on the number of conditions for PSPC, since the resulting equation $\psi^{(\rm i)\dagger}=\psi^{(\rm f)\dagger}\mathbf{V}$ can be directly obtained from $\psi^{(\rm f)}=\mathbf{V}\psi^{(\rm i)}$ by hermitean conjugation. In contrast, if $\psi^{(\rm f)}$ and $\psi^{(\rm i)}$ are orthogonal, an additional property is obtained, which leads to additional information as shown in the next section.

\subsection{Effective 2LS for orthogonal initial and final vectors}
\label{Eff_2ls}

Next we consider the simplest case of the $N$-level propagator, namely that all population resides in a single level prior to the pulse, $\psi^{{(\rm i)}}_k=\delta_{{\rm i}k}$, and also after the pulse, $\psi^{{(\rm f)}}_k=\exp(i\beta)\delta_{{\rm f}k}$. The dynamics in the present situation projects out column i of the propagator, $\psi^{(\rm f)}_j=\mathbf{V}_{j{\rm i}}$.

In case $N=2$ and $\rm i=1$ PSPI then leads to $\delta = \delta^{\rm PI}$ and $\Delta_2^{\rm PSPI} = -\frac{1}{2}\int_0^T {\rm tr}~\mathbf{H}(t') dt' - \beta + (2k+1)\pi$, $k\in\mathbb{Z}$, that is only 2 parameters are relevant due to orthogonal initial and final vectors. The PSPI propagator for $N=2$ reads
\begin{equation}
	\mathbf{V}^{(2),{\rm PSPI}} = \exp(i\chi)\left[
			\begin{array}{cc}
				0 										& e^{i\Delta_2^{\rm PSPI}} \\
				-e^{-i\Delta_2^{\rm PSPI}} & 0 									\\
			\end{array} \right]  \quad .
	\label{E_Prop_2LS_general}
\end{equation}
For $N$ levels this generalizes to $V_{{\rm f},k}=0$ for each $k\ne{\rm i}$. This follows from $V_{l,{\rm i}}=0$ for $l\ne {\rm f}$ and orthonormality of the $N$ column vectors of $\mathbf{V}$, which results in $0 = \sum_{l\ne{\rm f}} V_{l,k}V_{l,{\rm i}}^* = -V_{{\rm f},k}V_{{\rm f},{\rm i}}^*$ for all $k\ne {\rm i}$.

Orthogonal initial and final vectors correspond to the case $m = 4$, where the inversion property and $\mathbf{U}(T)\psi^{(\rm i)} = \psi^{(\rm f)}$ as well as $\mathbf{U}(-T)\psi^{(\rm i)} = -\psi^{(\rm f)}$ hold. Using $\mathbf{U}(-T)=\mathbf{V}^{\dagger}$, this case leads to the special condition $-\psi^{(\rm f)\dagger}=\psi^{(\rm i)\dagger}\mathbf{V}$. By the above argument, in row i of $\mathbf{V}$ all entries turn out to be zero except $V_{\rm i,f} = -\exp(-i\beta)$. Using the same reasoning as above we find that in column f only the ${\rm i}$-th element can be nonzero. This results in $2N-3$ additional (real) conditions, because $V_{\rm ii}=0$ has already been invoked above to enforce PSPI. 

For convenience we permute indices to get i$=1$ and f$=2$ such that the effective 2LS shows up in the PSPI propagator in terms of the block-diagonal form
\begin{equation}
	\mathbf{V}^{(N),{\rm PSPI}} = \left[
			\begin{array}{cc}
				\mathbf{V}^{(2),{\rm PSPI}}	& \mathbf{0} 		\\
				\mathbf{0} 						& \mathbf{V}^{(N-2)}	\\
			\end{array} \right]  \quad ,
	\label{E_Prop_Nlev}
\end{equation}   
where $\mathbf{V}^{(N-2)}$ is the propagator of the orthogonal subspace. It should be noted that $\mathbf{V}^{(N),{\rm PSPI}}\in {\rm U}(2)\oplus {\rm U}(N-2)\subset {\rm U}(N)$ {\it only} at the end of the pulse. The propagator is not block-diagonal while the pulse is on. 
Relating to the Floquet quasienergies and eigenvectors, we note that the $2\times 2$ block is equal to $\mathbf{V}^{(2),{\rm PSPI}}$, and hence eq.~(\ref{E_ResFloquetPhases}) indicates possible PSPI for orthogonal initial and final vectors independent of $N$.

In case of an initial superposition state we can use a rotation to transform $\psi^{(\rm i)}$, $\psi^{(\rm f)}$ to the first and second basis vector, respectively. In the rotated frame the propagator becomes  $\mathbf{V}_\mathbf{R}=\mathbf{R}\mathbf{V}\mathbf{R}^{\dagger}$. Defining $\psi_{\mathbf{R}}:=\mathbf{R}\psi$, the above derivation of the number of conditions extends to general initial states whenever $-\psi_{\mathbf{R}}^{(\rm f)\dagger}=\psi_{\mathbf{R}}^{(\rm i)\dagger}\mathbf{V}_{\mathbf{R}}$ holds. However, this clearly follows from $-\psi^{(\rm f)}=\mathbf{V}^{\dagger}\psi^{(\rm i)}$ under rotation. We note that $\mathbf{V}_\mathbf{R}^{\rm PSPI}$ will be block-diagonal but $\mathbf{V}^{\rm PSPI}$ will not in general assume such a form. 

We summarize that in the case of PSPI for arbitrary orthogonal wavefunctions an effective 2LS is obtained by fixing $4N-4$ real coefficients of the pulse. Thus the number of conditions may be larger than the $N^2-1$ parameters that determine the propagator. In this case we have to manipulate the {\it full} propagator to obtain an effective 2LS. Investigating $N=2$ as an example we have $4N-4=4$, while the propagator is determined by only $3$ parameters. The reason lies in the fact that when we demand $V_{11}=0$ and set the phase in $V_{12}$, the second column of $\mathbf{V}$ is already determined.
 
We note in addition that control is {\it independent} of the initial vector and only depends on the pulse, due to unitarity of the propagator and linearity of the Schrödinger equation. The propagator $\mathbf{V}$ therefore performs the same unitary transformation for every initial vector. 

If only populations matter we again have to replace $2N$ by $N$. The same arguments as mentioned above in this section lead to $N-1+N-2=2N-3$ conditions in case of a PI.

Whereas non-orthogonal initial and final vectors correspond to a control scenario rather than to population inversion, we can use similar arguments to estimate the number of parameters necessary for this extended control. In appendix \ref{App_nonortho} we show that for this setup the number of conditions increases to $8N-13$ (PSPC) and $8N-15$ (PC) for generating a 2LS, due to the absense of any symmetry. In appendix \ref{App_restrictions} we show furthermore that depending on the circumstances the number of conditions required may be smaller than derived above. The reason lies on the one hand in the upper bound of $N^2-1$ parameters for ${\rm SU}(N)$, which becomes essential for small enough $N$. On the other hand exceptional values of parameters can make other parameters redundant. An example is the irrelevance of $\Delta_1$ when $\cos(\delta)=0$ in case of $N=2$. 

\subsection{The adiabatic limit}

Previously Holthaus \cite{C_holthaus_prl92,C_holthaus_pra94} proposed the following relation among the instantaneous Floquet quasienergies in case of PI by a pulse with adiabatic envelope,
\begin{equation}
	\int_0^T dt [\mathbf{\Omega}^{({\rm adiab})}_{\rm f}(t)-\mathbf{\Omega}^{({\rm adiab})}_{\rm i}(t)] = (2n+1)\pi \quad , \quad n\in\mathbb{Z} 		
		\quad ,
	\label{E_adiabRes}
\end{equation}
by replacing exact energy eigenstates (no field) by integrals over time-dependent quasienergies in phase-factors, using Floquet theory for continuous waves \cite{C_holthaus_prl92}. In a later publication, RWA on resonance (zero detuning) is compared with Floquet theory and the relation between quasienergies calculated for this special case is used to conjecture that eq.~(\ref{E_adiabRes}) remains valid also for RWA with nonzero detuning \cite{C_holthaus_pra94}.
In the following we show how eq.~(\ref{E_adiabRes}) can be rigorously deduced from eq.~(\ref{E_ResFloquetPhases}) whenever the envelope can be treated adiabatically.

Assume a pulse of duration $T$, represented by the product of a periodic function and a slowly varying envelope. Note that in general the periods can even have different length, $d_p$, with $p$ labelling the period, if attached e.g. at zeros to give a continuous field. Due to slow variation of the envelope we make the assumption that the envelope is constant within each period $p$. We can represent the propagator $\mathbf{V}$ as a product of contributions over all periods, 
\begin{equation}
	\mathbf{V} = \prod_{p=1}^{P} \Theta_p e^{-i\mathbf{\Omega}^{(p)}d_p} \Theta_p^{\dagger}
\end{equation}
using standard Floquet theory within each of the $P$ periods. $\mathbf{\Omega}^{(p)}$ denotes the matrix of Floquet quasienergies for period $p$ and $\Theta_p$ transforms to the diagonal frame. Making the well-known additional (adiabatic) approximation that $\Theta_p^{\dagger}\Theta_{p-1}\approx \mathbf{I}_N$ for every $p$, we arrive at $\mathbf{V} \approx \Theta_P \exp\left(-i\sum_{p=1}^{P}\mathbf{\Omega}^{(p)}d_p\right)\Theta_1^{\dagger}$. This has to be compared to $\Theta e^{-i\mathbf{\Omega} T} \Theta^{\dagger}$, with $\mathbf{\Omega}$ and $\Theta$ corresponding to the whole pulse (time interval $[0,T]$), within our approach of periodically repeating the pulse. Note $\Theta_p^{\dagger}\Theta_{p-1}\approx \mathbf{I}_N$ applied to every $p$ means that the envelope was actually replaced by a rectangle of duration $T$. Therefore we get $\Theta=\Theta_p$, for every $p$, and after finally using that $\Theta_p^{\dagger}\Theta_{p-1}\approx \mathbf{I}_N$ implies that $\mathbf{\Omega}^{(p)}-\mathbf{\Omega}^{(p-1)}$ is very small, we arrive at
\begin{equation}
	\mathbf{\Omega}  = \frac{1}{T}\sum_{p=1}^{P}\mathbf{\Omega}^{(p)}d_p \approx \frac{1}{T}\int_0^T dt \mathbf{\Omega}^{({\rm adiab})}(t) 
		\quad .
\end{equation}

Together with eq.~(\ref{E_ResFloquetPhases}) this leads immediately to eq.~(\ref{E_adiabRes}).
We note that in case of a pulse with slowly varying envelope and all $d_p$ equal (i.e., a well defined carrier frequency), eq.~(\ref{E_adiabRes}) yields extra information on the transient dynamics. The cost, however, are the restrictions imposed by the adiabatic approximation.  

\section{Summary}

In the present paper we investigate control of the full final wavefunction (phase-selective population control, PSPC) or just final population control (PC) within the semiclassical dipole approximation. The analysis is valid for any pulse having finite support (of length $T$) in time if the equation of motion is local in time (no or only adiabatic memory effects).

For orthogonal initial and final vector, population inversion (PI) in the 2-level system is governed by a single parameter (here denoted $\delta$). In terms of Floquet quantities we obtain the criterion $(\omega_{\rm f}-\omega_{\rm i})T=(2n+1)\pi$, $n\in\mathbb{Z}$ on the difference of Floquet quasienergies from the 2d subblock of the propagator corresponding to the initial and final vector. This criterion is however only necessary, as it can be fulfilled by varying another parameter of the propagator not leading to PI. A sufficient criterion is provided by the shape of the Floquet eigenvector matrix. In contrast, for phase-selective population inversion (PSPI) $2$ parameters of the propagator have to be set appropriately.
Applying the adiabatic approximation we show that our criterion specializes to the expression presented by Holthaus et.~al.~\cite{C_holthaus_prl92,C_holthaus_pra94}.

For $N$-level systems we find that for PSPC at most $2N$ real parameters of the propagator have to be set, while $N$ conditions are sufficient for PC. The propagator at $t=nT$, $n\in\mathbb{N}$, can always be made block-diagonal with one subblock of dimension $2\times 2$, that is, initial and final level form a $2$d subspace, decoupled from the $(N-2)$-dimensional subspace of the remaining levels. 
In order to create such an effective 2LS in a given basis with preselected initial and final vectors, additional conditions must be fulfilled; for PSPI at most $4N-3$ real parameters must be fixed while for PI $2N-2$ parameters are needed and moreover the criterion on the Floquet quasienergies derived for $N=2$ holds true unaltered for any $N>2$. 
In case of PSPC for non-orthogonal initial and final vectors we need to set $8N-12$ real parameters to obtain the block-diagonal form with $2\times 2$ subblock, while it is $8N-14$ for PC. The numbers of conditions derived are in general upper bounds due to possible dependencies between parameters of ${\rm SU}(N)$.

\appendix

\section{Effective 2LS for non-orthogonal initial and final vectors}
\label{App_nonortho}
 
We now treat the case that $\psi^{(\rm i)}$ and $\psi^{(\rm f)}$ are two non-orthogonal vectors of length $1$, for which we derive the number of conditions necessary to obtain an effective 2LS. Remembering that the Floquet quasienergies can be written as $\omega_j=q_j\omega_0$, $\omega_0:=2\pi/T$ and $q_j=n_j/m\in[0,1)$, we have in the present case $m\ne 4$ or at least one of the $q_j$ will even be irrational. The dynamics does not involve only 2 but an arbitrary number of all $N$ levels. However, as all points of the trajectory are located on the great circle,
it should be possible to transform the propagator to block diagonal form, with a $2\times 2$ block $\mathbf{V}_\mathbf{R}^{(2)}$ corresponding to the subspace spanned by the initial and final vectors in a rotated frame. We first construct a basis consisting of $\psi^{(\rm i)}$, a second vector orthogonal to $\psi^{(\rm i)}$ and coplanar with $\psi^{(\rm i)}$ and $\psi^{(\rm f)}$, and $N-2$ vectors orthogonal to both $\psi^{(\rm i)}$ and $\psi^{(\rm f)}$. A rotation $\mathbf{R}$ relates the standard basis to this new one and the propagator becomes $\mathbf{V}_\mathbf{R}=\mathbf{R}\mathbf{V}\mathbf{R}^{\dagger}$. Such a transformation always exists, and in the transformed basis at most the first and second levels are populated in the initial and final state. 

For control it remains to manipulate $\mathbf{V}_\mathbf{R}^{(2)}$. In contrast to orthogonal $\psi^{(\rm i)}$ and $\psi^{(\rm f)}$ this requires setting all (complex) elements of $\mathbf{V}_{\mathbf{R};jk}$, $\{j<3\}\wedge \{k>2\}$ or $\{k<3\}\wedge \{j>2\}$, to zero. Adding in case of PSPC the $3$ parameters of $\mathbf{V}_\mathbf{R}^{(2)}$, we have to fix a total of $2\times 2\times 2(N-2) + 3 = 8N-13$ real parameters to obtain an effective 2LS. The larger number of parameters is required due to the lack of any symmetry. 

In case of PC only one parameter, $\delta$ of $\mathbf{V}_\mathbf{R}^{(2)}$, is relevant. However we also have to set the $8N-16$ appropriate elements of $\mathbf{V}_\mathbf{R}$ to zero, and thus we still need to impose $8N-15$ conditions.          

\section{Restrictions on the number of conditions}
\label{App_restrictions}

As noted in subsection~\ref{Eff_2ls}, the number of conditions derived for control and for creating an effective 2LS may exceed the total number of parameters of the propagator. Noting that the multiplicative phase $\chi$ in eq.~(\ref{E_Prop}) is predetermined by ${\rm tr}(\mathbf{H})$, we require $N^2-1$ parameters specifying the propagator. The number of conditions derived in subsection~\ref{Eff_2ls} and appendix~\ref{App_nonortho} is relevant only if it is {\it smaller} than $N^2-1$, which in case of PSPC leads to $N > 2$, and for an effective 2LS to $N > 3$ (PSPI) and $N > 6$ (PSPC). 

In order to solve the homogeneous system of equations from the Floquet matrix (which in general is infinite), the Floquet determinant has to be set to zero \cite{C_Floquet}. This adds one real condition in case of a hermitean Hamiltonian. 

Assuming $N>1$ this leaves in total $2N$ real parameters for PSPC, and $\min(4N-3,N^2)$ or $\min(8N-12,N^2)$ real parameters of the pulse to obtain an effective 2LS for orthogonal and nonorthogonal vectors, respectively. The total number of conditions are $N$ for only PC, and $2N-2$ or $\min(8N-14,N^2)$ to obtain an effective 2LS for PC with orthogonal and nonorthogonal vectors, respectively.
For better overview we present results for $N\le 10$ in table \ref{table_conditions}.

There is a further possible reduction of the number of parameters for necessary for control. For an illustration we come back to $N=2$ and note that for orthogonal initial and final vector the parameter $\Delta_1$ was irrelevant in case $\cos(\delta)=0$. This situation of pure states (only one level of $\psi^{(\rm i)}, \psi^{(\rm f)}$ populated in the appropriate basis) restricts to a subset of ${\rm SU}(2)$. For $N>2$ we expect such situations to occur as well, such that our numbers of conditions become upper bounds in general. Although a general parametrization of ${\rm SU(}N)$ via a faithful matrix representation of its algebra with minimal dimension has been given \cite{C_Tilma2002}, more work is necessary to find detailed corrections to the number of conditions.



\newpage

\begin{table}
	\begin{tabular}{|c||c|c|c|c|c|c|}
		\hline
		$N$ & \multicolumn{2}{|c|}{\textbf{Control}} & \multicolumn{4}{|c|}{\textbf{Effective 2LS}} \\
				& \multicolumn{2}{|c|}{}					&	\multicolumn{2}{|c|}{orthogonal $\psi^{(\rm i)}$,$\psi^{(\rm f)}$} & 		
			\multicolumn{2}{|c|}{non-orthogonal $\psi^{(\rm i)}$,$\psi^{(\rm f)}$} \\
		\cline{2-7}
				&	PSPC	&	PC	&	PSPI	&	PI	&	PSPC	&	PC	\\
		\hline\hline
		\bf 2		&	4		& 2		& 3 	&	2		&	4		&	2		\\
		\hline
		\bf 3		&	6		& 3		& 9		&	4		&	9		&	9		\\
		\hline
		\bf 4		&	8		& 4		& 13	&	6		&	16	&	16	\\
		\hline
		\bf 5		&	10	& 5		& 17	&	8		&	25	&	25	\\
		\hline
		\bf 6		&	12	& 6		& 21	&	10	&	36	&	34	\\
		\hline
		\bf 7		&	14	& 7		& 25	&	12	&	44	&	42	\\
		\hline
		\bf 8		&	16	& 8		& 29	&	14	&	52	&	50	\\
		\hline
		\bf 9		&	18	& 9		& 33	&	16	&	60	&	58	\\
		\hline
		\bf 10	&	20	& 10	& 37	&	18	&	68	&	66	\\
		\hline		
	\end{tabular}
	\caption{Number of conditions as a function of the number of levels $N$ for control and for
		generating an effective 2-level system for orthogonal and non-orthogonal initial $\psi^{(\rm i)}$ 
		and final $\psi^{(\rm f)}$ vector, respectively. Note due to depedencies of parameters of $SU(N)$ 
		the necessary conditions may be less than given here for $N>2$.
	}
	\label{table_conditions}
\end{table}

\end{document}